\documentclass[aps, showpacs, epsf, twocolumn]{revtex4}
\usepackage{graphicx}
\usepackage{dcolumn}
\usepackage{amsmath}
\usepackage{amssymb}
\usepackage{ulem}
\usepackage[colorlinks]{hyperref}
\hypersetup{citecolor=blue}

\usepackage{bm}
\usepackage{epstopdf}
\usepackage{amsthm}
\usepackage{float}
\graphicspath{{./figures/}}

\def\epl#1#2#3{{Europhys. Lett.} {\bf #1}, #2 (#3)}
\def\ijbc#1#2#3{{Int. J. Bifurcation Chaos} {\bf #1}, #2 (#3)}

\def\prl#1#2#3{{ Phys. Rev. Lett.} {\bf #1}, #2 (#3)}

\def\pre#1#2#3{{ Phys. Rev. E} {\bf #1}, #2 (#3)}

\def\chaos#1#2#3{{Chaos} {\bf #1}, #2 (#3)}

\def\prl#1#2#3{{ Phys. Rev. Lett.} {\bf #1}, #2 (#3)}

\def\pre#1#2#3{{ Phys. Rev. E} {\bf #1}, #2 (#3)}

\def\ie{i.e., }

\def\ep{\varepsilon}

\def\ie{i.e.}

\def\etal{{\it et al.}}

\def\beqr{\begin{eqnarray}}
\def\eqnr{\end{eqnarray}}
\def\beq{\begin{equation}}
\def\bc{\begin{center}}
\def\ec{\end{center}}
\def\eqn{\end{equation}\noindent}
\begin{document}

\title{Collective Dynamics of coupled Lorenz oscillators near the Hopf Boundary: Intermittency and Chimera states}

\author{Anjuman Ara Khatun$^1$, Yusra Ahmed Saeed$^{1,2}$, Nirmal Punetha$^3$, and Haider Hasan Jafri$^1$}

\affiliation{$^1$Department of Physics, Aligarh Muslim University, Aligarh 202 002, India\\
$^2$Physics Department, Taiz University, Yemen,\\
$^3$Amity University Haryana, Gurgaon 122413 India}

\begin{abstract}
We study collective dynamics of networks of mutually coupled identical Lorenz oscillators near subcritical Hopf bifurcation. 
This system shows induced multistable behavior with interesting spatio-temporal dynamics including synchronization, desynchronization and chimera states.  
We find this network may exhibit intermittent behavior due to the complex basin structures, where, temporal dynamics of the oscillators in the ensemble switches between different attractors. 
Consequently, different oscillators may show dynamics that is intermittently synchronized (or desynchronized), giving rise to {\it intermittent chimera states}. 
The behaviour of the intermittent laminar phases is characterized by the characteristic time spend in the synchronization manifold, which decays as power law.
This intermittent dynamics is quite general and can be extended for large number of oscillators interacting with nonlocal, global and local coupling schemes.
\end{abstract}
\maketitle

%%%%%%%%%%%%%%%%%%%%%%%%%%%%%%%%%%%%%%%%%%%%%%%%%%%%%%%%%%
%%%%%%%%%%%%%%%%%%%%%%%%%%%%%%%%%%%%%%%%%%%%%%%%%%%%%%%%%%
\section{Introduction} \label{sec:intro}

For many decades now, the Lorenz equations \cite{Lorenz}, 
\begin{equation} \label{eq:system-eq0}
\begin{split}
\dot{x} = \rho(y-x);~
\dot{y} = \gamma x - y - xz;~ 
\dot{z} = xy - \beta z,
\end{split}
\end{equation}
used for describing convection rolls in the atmosphere, has served as paradigmatic model to study dynamical properties of chaotic systems. Typical parameter values considered in such studies~\cite{Sparrow-1982}, \ie, $\rho=10, \gamma=28$ and $\beta=8/3$, ensure asymptotic motion on a well known Lorenz chaotic attractor. This dynamically rich model has been used in number of studies as a standard prototype for exploring properties of chaotic dynamics~\cite{Sparrow-1982} and phenomena such as chaos synchronization~\cite{Pecora-1990, Pecora-1997, Pecora-2015, Camargo-2012}, amplitude death \cite{konishi-pre-2004, karnatak-pre-2007}, chimera dynamics \cite{ujjwal-pre-2017,  shepelev-chaos-2018, xiao-nd-2018} and intermittency \cite{shepelev-chaos-2018, koronovskii-pre-2020}.  

The idea that chaotic systems can be driven to synchrony was introduced by Pecora and Carroll \cite{Pecora-1990}. Since then, different scenarios of synchronization have been extensively studied in a wide range of periodic and chaotic systems with various coupling strategies\cite{Lai-1993, Karnatak-2009, Murali-1994}. These include generalized synchrony~\cite{Rulkov-1995}, complete~\cite{Pecora-1990,Fujisaka-1983}, phase and lag synchrony~\cite{Rosenblum-1997}. The study of the phenomenon of synchronization, due to its ubiquitous presence in wide range of real world systems, has found applications in various fields including physics, biology, social science and engineering~\cite{sync-book}. 

A related phenomenon which is observed close to the boundary of chaotic synchrony is intermittency. This interesting dynamics refers to a situation where phase synchronized oscillations (laminar phases) are interrupted by nonsynchronous behaviour (phase slips) during persistent time intervals. Intermittency is observed close to the threshold parameter values for which the coupled system is synchronized. Various studies have explored its origin and statistical properties and different type of intermittent behaviors have been classified as type -I, -II \cite{Berge-1988, Dubois-1983} or on-off intermittency \cite{Platt-1993,Heagy-1994}. These pre-transitional intermittencies have been characterized near lag \cite{Rosenblum-1997a, Boccaletti-2000,Zhan-2002}, generalized \cite{Hramov-2005}, and phase \cite{Pikovsky-1997, Lee-1998, Boccaletti-2002,Rosa-1998} synchronization regime. The statistics of distributions of Lyapunov exponents have been applied to the case of intermittency~\cite{Prasad-1999, Dutta-2003}. Studies have shown that switching of dynamics between two or more distinct types of behaviour reflects in the distribution of largest Lyapunov exponents (LLEs)~\cite{Prasad-1999}.

An interesting spatio-temporal pattern called Chimera state is another collective behavior observed in ensemble of oscillators. In this state, the oscillator population spontaneously splits into synchronized and desynchronized populations breaking its symmetry~\cite{Kuramoto-2002, Abrams-2004, Panaggio-2015}. Recently, these states have gained high interest due to its interesting spatio-temporal nature with two dynamically distinct properties, namely, synchronized and desynchronized motions, coexisting in a single population. These mixed states have applications in understanding the phenomena such as ventricular fibrillation, unihemispheric and REM (rapid eye movement) sleep, power grid stability and consensus formations in social networks~\cite{Panaggio-2015}. 

Chimera states are observed in several oscillator ensembles from large oscillators networks where thermodynamic limit can be applied~\cite{Kuramoto-2002, Abrams-2004}, to a network as small as a system of three oscillators~\cite{Ashwin-2015,Panaggio-2016,Bick-2016,Bick-2017,Bohm-2016,Hart-2016,Dudowski-2016,Wojewoda-2016,Maistrenko-2017}. In general, chimera states in oscillator ensembles are known to emerge when some degree of non-uniformity is introduced in the system~\cite{Panaggio-2015}, for example, in the form of nonlocal interactions~\cite{Kuramoto-2002, Abrams-2004}, modularity~\cite{Chandrasekar-2014, Gopal-2014, Punetha-2015, Ujjwal-2016-a}, parameter heterogeneities~\cite{Sakaguchi-2006,Laing-2010}, time-delay couplings~\cite{Sethia-2008} or amplitude fluctuations ~\cite{Sethia-2014}. However, recent studies suggest that even in the absence of such non-uniformities, in a globally coupled ensemble of identical chaotic oscillators, chimeric states can be generated~\cite{Ujjwal-2016-b, Ujjwal-2017} and controlled~\cite{Anjuman-2021, Anjuman-pre-2021} through induced multistability. In this case, interactions between the oscillators modify effective control parameters of the system, thus shifting it towards a multistable regime where attractors with contrasting synchronization properties coexist. In this multistable regime, depending on the initial conditions, the oscillators from the population can settle into any of these coexisting attractors with dynamically different properties. Since both synchronized and desynchronized motions are possible on these attractors, the system exhibits chimeric behavior. The basins of attraction in such systems were found to be interwoven and riddled~\cite{Ujjwal-2016-b} with multiplicity of coexisting attractors~\cite {Wontchui-2017}. 

In this work, we consider networks of coupled Lorenz oscillators on a ring and obtain chimera states in the system through induced multistability. We examine temporal behavior of the oscillators in such ensemble and observed intermittent dynamics: the trajectories of the oscillators may jump from one attractor to another leading the system to show {\it intermittent chimeras}. We found that synchronization and desynchronization manifolds in the system, due to its complex structure, may push or pull oscillators' trajectories towards (or away from) a particular attractor resulting in such intermittent behavior. This intermittent dynamics is characterized with the help of local divergence plots \cite{gao-1993,gao-1994}. We numerically observe that time-interval of laminar phases scales as a power law. These results are verified for the networks of different sizes, $N = 4, 7, 10$ and $100$. We also extend our study for more general scenario of nonlocally coupled oscillator networks.%

The organization of the manuscript is as following. In the next section \ref{sec:lorenz-ring}, we consider Lorenz oscillators on a ring topology. We show multiple attractors observed in example system of three oscillators when multistability is induced in such ensembles through the coupling. We discuss the dynamical scenarios that are observed on these attractors and how this may lead to an intermittent behavior including the occurrence of `intermittent chimera states'. This phenomenon is examined by looking at the time-evolution of the system and finite-time Lyapunov exponent. We analyze intermittent dynamics using local exponential divergence plots and interleaved time-intervals. We also verify and discuss the generality of these results for different larger size networks. In Sec.~\ref{sec:nonlocal}, we further extend our study to network of nonlocally coupled oscillator. This is followed by a summary and discussion in section \ref{sec:conclusion}.

%%%%%%%%%%%%%%%%%%%%%%%%%%%%%%%%%%%%%%%%%%%%%%%%%%%%%%%%%%
%%%%%%%%%%%%%%%%%%%%%%%%%%%%%%%%%%%%%%%%%%%%%%%%%%%%%%%%%%
\section{Lorenz oscillators on a ring}

\label{sec:lorenzring} \label{sec:lorenz-ring}
We consider a ring of $N=3$ coupled Lorenz oscillators described mathematically by the following equations, 
\begin{equation} \label{eq:system-eq1}
\begin{split}
\dot{x}_{i} &= \rho(y_{i}-x_{i}), \\ 
\dot{y}_{i} &= \gamma x_{i} - y_{i} - x_{i}z_{i}, \\
\dot{z}_i &= x_i y_i - \beta z_i + \varepsilon \bigtriangleup {z_i},
\end{split}
\end{equation}
\noindent
where $\bigtriangleup {z_i} = z_{i-1} - 2z_i + z_{i+1}$, the index $i = 1,2, \cdots N$ and $i$ is taken module $N$. Oscillators are connected through $z_i$-variables. Here $\ep$ represents the  coupling strength which is introduced in such a way that each node is connected to the left and to the right with equal strength. Since Lorenz system is invariant under the transformation $(-x_i, -y_i, z_i) \rightarrow  (x_i, y_i, z_i)$,  coupling through $z_i$ variables preserves the symmetry of the system in the $x_i, y_i$  planes~\cite{Ujjwal-2016-b}. The parameter values are taken $\rho=10$, $\gamma=24.76$ and $\beta=8/3$, such that the fixed points of the isolated oscillators $(\pm{\sqrt{ \beta(\gamma-1)}},\pm{\sqrt{ \beta(\gamma-1)}},\gamma-1)$~\cite{Sparrow-1982, Strogatz-Book-1994} are unstable and the dynamics of each of the uncoupled oscillator is chaotic. 

% Figure ------------------------------------------------------------------
\begin{figure} 
\includegraphics [scale=0.40,angle=0]{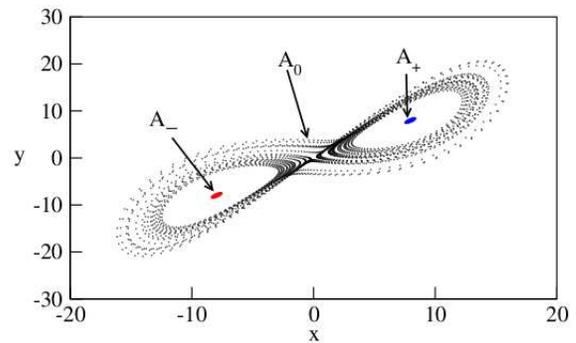}
\caption{Multistable attractors $A_{-}, A_{+}, A_{0}$ observed in a coupled Lorenz oscillators (Eq.~\ref{eq:system-eq1} ) on a ring at coupling strength $\ep = 0.02$.}
\label{fig1}
\end{figure}

% Figure ------------------------------------------------------------------
\begin{figure} 
\includegraphics [scale=0.40,angle=0]{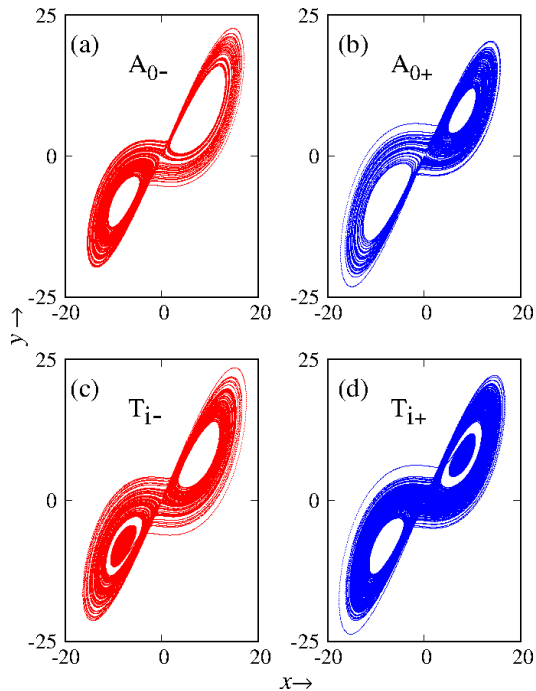}
\caption{(Color online) Different dynamical behavior observed in the system. At $\ep = 0.05$, the biased motion is shown in (a) and (b), where the trajectory remains in one of the lobes while moving on attractor $A_0$, denoted by $A_{0+}$ and ${A_0}_-$. For higher coupling ($\ep = 0.10$), the trajectories on an attractor can also jump to another attractor as shown in (c) and (d), denoted by trajectories ${T_i}_+$ and $T_{i-}$.}
\label{fig2}
\end{figure}

In the absence of coupling, the oscillators in this ensemble settle into the well known Lorenz chaotic attractor $A_0$. In coupled system, however, one observes two new symmetric stable attractors denoted by $A_+$ and $A_-$ along with the typical attractor $A_0$ (See Fig.~\ref{fig1}). It is observed that the oscillators settling into attractors $A_{+}$ and $A_{-}$ show synchronized behavior while the motion on $A_0$ is desynchronized~\cite{Ujjwal-2017}. These distinct synchronization properties of the attractors give rise to induced chimera states through mutistability. In the system of $N$ oscillators, the dynamics can take place in three possible manifolds. A fully desynchronized motion on $A_0$ takes place in a $3N$-dimensional desynchronized manifold $\mathcal{M}_d$. Whereas, when synchronization is established in the system, the condition $x_i = x_j, y_i = y_j, z_i = z_j$, for all $i,j$ is satisfied. These $(3N-3)$ constraints on the motion reduce dimensionality of the system constituting a three dimensional subspace, which is, synchronization manifold, $\mathcal{M}_{s}$. Similarly, fulfilment of anti-synchronization condition, $x_i = -x_j, y_i = -y_j, z_i = z_j$ represent a motion on anti-synchronization manifold $\mathcal{M}_{a}$. Dimensions of these subspaces depend upon the number of constraints, \ie, number of synchronized and anti-synchronized oscillators. Coexistence of desynchronized manifold with either one, or both of these other manifolds implies the existence of chimeric states. 

The typical attractors observed in the system of coupled Lorenz system are shown in Fig.~\ref{fig1}. Depending on the initial condition, the oscillators in the ensemble can end up in any of these attractors. As mentioned before, these coexisting attractors are responsible for system's chimeric behavior due to their distinct synchronization properties: oscillators settling into attractor $A_0$ show desynchronized behavior however the ones which asymptote to $A_{+}$ and $A_{-}$ show synchrony while being in anti-synchrony with each other. Besides these attractors, additional dynamical motions which we observe on a ring of coupled Lorenz oscillators are shown In Fig.~\ref{fig2}. For example, a biased motion in shown in Fig.~\ref{fig2} (a, b), where the trajectory spends more time in one of the wings of the chaotic Lorenz attractor $A_0$. Further, we observe that oscillators may not be contained within one attractor, and can jump from one attractor to another as shown in Fig.~\ref{fig2} (c, d), where the trajectory jumps between attractors, represented by trajectories $T_i$. 

\begin{figure}
\centering
\includegraphics[scale=0.45,angle=0]{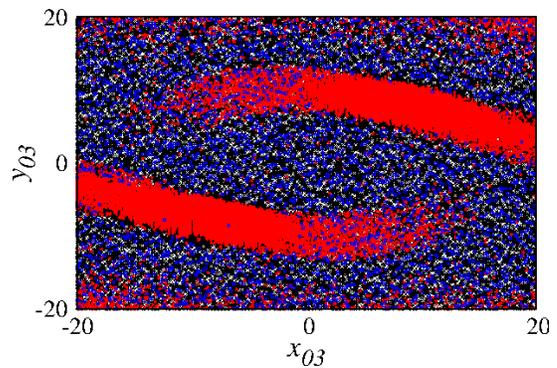}
\caption{(Color online) Basin of attraction of the system of three coupled Lorenz oscillators on a ring at coupling strengths $\varepsilon = 0.028$. Basins are calculated for initial conditions in $x_3$ and $y_3$ variables, i.e., $x_{03}$, $y_{03}$ space in the interval $[-20, 20]$. Initial values of $x_1, y_1, z_1, x_2, y_2, z_2, z_3$ are fixed, i.e., $x_{01} = x_{02} = y_{01} = y_{02} = z_{01} = z_{02} = z_{03} =1$. The initial conditions evolving towards different dynamics are represented by white, blue (dark grey), red (light grey) and black regions (see text for details).}
\label{fig3} 
\end{figure}

Globally coupled Lorenz oscillators with multistable behavior have complex basin structures~\cite{Camargo-2012}, giving rise to the chimera states~\cite{Ujjwal-2017} in such systems. Similar to globally coupled oscillators, the basin of attraction of this system also shows riddled behavior: trajectories starting with infinitesimally close initial conditions have a nonzero probability of ending up into different attractors. The complexity of the basin implies that the manifolds containing oscillator trajectories are also very complex. 
In Fig.~\ref{fig3}, we plot the basin of attraction of the system of three coupled oscillators at $\ep = 0.028$. Here, white region are where all the oscillators evolve to $A_0$ attractor and the dynamical state can be represented as ($A_0, A_0, A_0$). Red (light grey) and blue (dark grey) region correspond to the collective states ($A_0,A_{\pm},A_{\pm}$) and ($A_0, A_0, A_{\pm}$) respectively. Black region indicates a collective state $T_{i\pm}$ where at least one oscillator has intermittent dynamics and the other oscillators may evolve to any of the coexisting attractors.

% Figure ------------------------------------------------------------------

% Figure ------------------------------------------------------------------
\begin{figure}
\centering
\includegraphics[scale=0.37,angle=0]{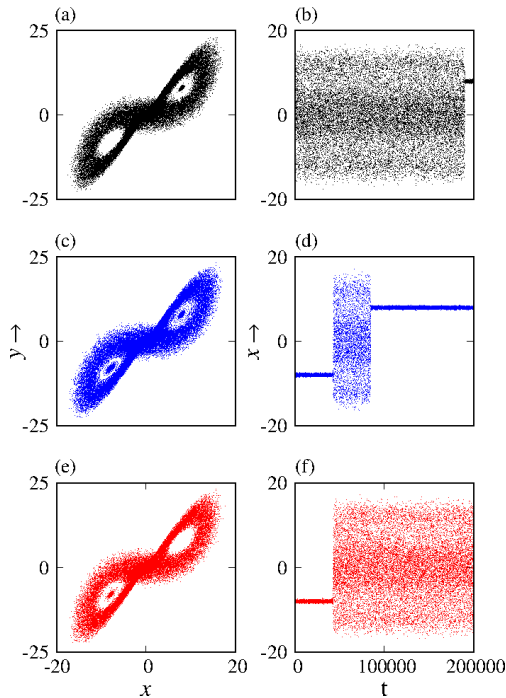}
\caption{(Color online) Plots for intermittent dynamics (left panel) and their corresponding time series (right panel). In (a) oscillator spends large amount of time in $A_0$ attractor but jumps to $A_{+}$. (b) represents their corresponding time series. In (c) and (d), we observe two jumps from initial attractor $A_-$ to $A_0$ and then to $A_+$. A jump from smaller attractor $A_-$ to bigger attractor $A_0$ is shown in (e) and (f).}
\label{fig4} 
\end{figure}

% Figure ------------------------------------------------------------------
\begin{figure}
\centering
\includegraphics[scale=0.40,angle=0]{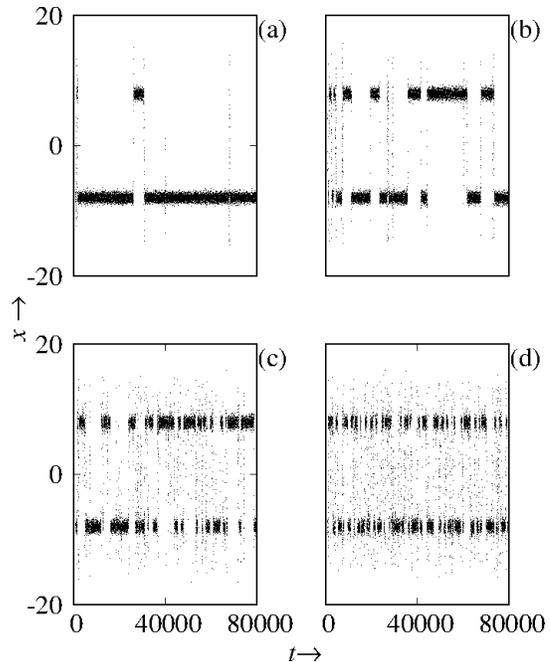}
\caption{Effect on intermittent behavior for different coupling strength values. Time series of variable $x$ is plotted to see intermittent dynamics at (a) $\ep=0.08$, (b) $\ep=0.09$, (c) $\ep=0.11$ and (d) $\ep=0.12$.}  
\label{fig5} 
\end{figure}

The intermittent dynamics of the oscillators observed for the coupled system at $\ep=0.03$ is shown in Figs.~\ref{fig4}. The first oscillator spends large amount of time in $A_0$ type state, but then jumps to $A_+$ attractor as shown in Fig.~\ref{fig4} (a) and (b). Dynamics of the second oscillator is initially on $A_-$ followed by $A_0$ and jumps to $A_+$ type attractor. This is shown in Figs.~\ref{fig4}(c) and (d). As shown in Figs.~\ref{fig4}(e) and (f), the dynamics of the third oscillator is on $A_-$ attractor before jumping to $A_0$. For all these cases system may again jump to another attractors when it is evolved further. Time series plots of the $x$ variable are plotted in Fig.~\ref{fig5} to show more prominent intermittent dynamics at higher coupling strengths. The hopping between different dynamical states becomes fast as the value of coupling strength is increased~\cite{Grebogi} as shown in Fig.~\ref{fig5}(b)-(d).

% Figure ------------------------------------------------------------------
\begin{figure}
\centering
\includegraphics[scale=0.35,angle=270]{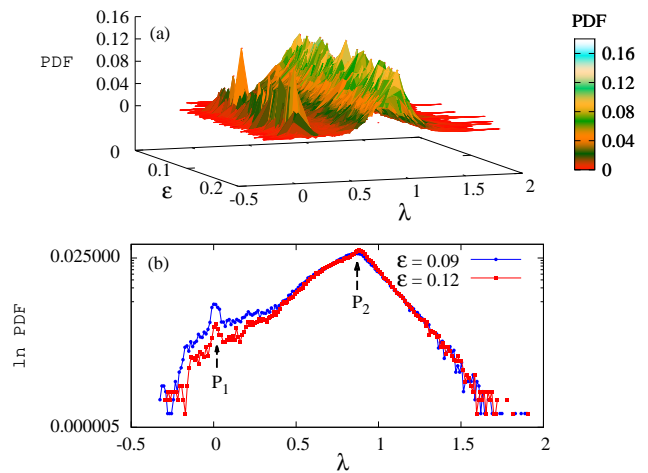}
\caption{(Color online) The distributions (PDF) of finite time Lyapunov exponents $\lambda$ are plotted with coupling strength $\varepsilon$. Two visible peaks in the figure correspond to two different attractors and indicate that the trajectory may switch between two dynamical behaviors in time. In (b) we plot the logarithm of the PDF showing that the two peaks can be interpolated through straight line in the semi-log graph. These exponents are calculated along a very long trajectory which is divided into segments of equal finite-time lengths $t = 10$. The finite-time Lyapunov exponent is then calculated for each segment. 
}  
\label{fig6} 
\end{figure}

% Figure ------------------------------------------------------------------
\begin{figure}
\centering
\includegraphics[scale=0.33,angle=270]{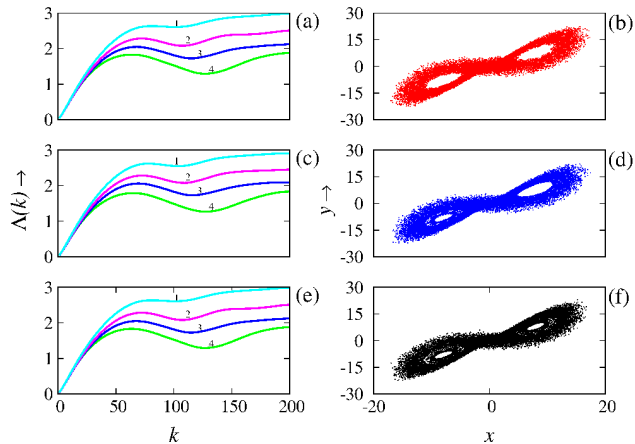}
\caption{(Color online) The $\Lambda (k)$ curve for the time series of an intermittent attractor  obtained at $\ep =0.09$. The numbers $1$ to $4$ corresponds to shells defined by $(2^{-(i+1)/2},  2^{-i/2})$ with $i=$9 to 12.}  
\label{fig7} 
\end{figure}

This complex intermittent behavior can be verified by plotting the distribution of finite-time Lyapunov exponents $\lambda$. With increasing strength of coupling, we examine finite-time Lyapunov exponents calculated for relatively small time-intervals taken from a long trajectory, and plot its probability distribution function (PDF) with coupling strength in Fig.~\ref{fig6}. Finite-time Lyapunov exponents provide information about average convergence or divergence rate of two nearby trajectories within the attractor for a given time-window. Since here the trajectories may switch between different attractors, two peaks are observed due to different convergence (or divergence) rate on those attractors. In Ref.~\cite{Prasad-1999}, it was shown that in case of intermittency, the PDF of the local LE is a combination of a normal density and a stretch exponential tail. For the case of an uncoupled map, authors have shown that for intermittency, the local LE has a Gaussian distribution centered at different values and a stretched exponential tail that interpolates between the two. The distribution of the exponents indicates that the trajectories in some time-windows can fall apart or can come close while moving within an attractor. Two peaks in the PDF correspond to the states between which intermittent dynamics takes place. In the present case, the PDFs of the local LE are plotted in Fig.~\ref{fig6}(a) for different values of $\ep$. As shown, the PDF consists of two peaks $P_1$ and $P_2$ which clearly indicates presence of two distinct behaviours. To understand the interpolation between the two peaks ($P_1$ and $P_2$)~\cite{Prasad-1999}, we plot the PDF in Fig.~\ref{fig6}(b) for two values of $\ep$ and observe that it is a straight line indicative of an exponential behaviour.

We characterize these intermittent states by using local exponential divergence plots proposed by Gao {\etal} \cite{gao-1993,gao-1994} as following. These plots are used to characterize complex motions and can be applied directly to the experimental time series. Here we outline the procedure to calculate this exponent for the time series given by $x(1), x(2), \dots, x(N)$ normalized to the unit interval $[0,1]$. We then make use of time delay embedding \cite{Takens-1981, Sauer-1991} procedure to construct vectors of the form $X_i=[x(i),x(i+L),\dots,x(i+(m-1)L)]$, where $m$ is the embedding dimension and $L$ is the delay time to be chosen such that the optimization criteria is satisfied. The time dependent exponent $\Lambda(k)$ is defined as 
\beq
\Lambda(k)= \left \langle \ln \left( \frac{||X_{i+k}- X_{j+k}||}{||X_i-X_j||} \right) \right \rangle
\eqn

where $d \le ||X_i-X_j|| \le d+\Delta d$, with $d$ and $\Delta d$ being small distances and $||.||$ denotes a Euclidean norm. Angular bracket denotes ensemble average over all possible pairs $(X_i,X_j)$ and $k$ is the evolution time. Computation is carried out for a series of shells $(d,d+\Delta d)$. For a purely chaotic signal, the $\Lambda (k)$ curve first increases linearly and then flattens. Linear regime for different shells collapses to form a common envelope. As shown in Fig.~\ref{fig7}, we plot $\Lambda(k)$ versus $k$ curve for the intermittent dynamics at $\ep=0.09$ and we note that the common envelope is well defined indicating that the dynamics is chaotic.

% Figure ------------------------------------------------------------------
\begin{figure}
\centering
\includegraphics[scale=0.32,angle=270]{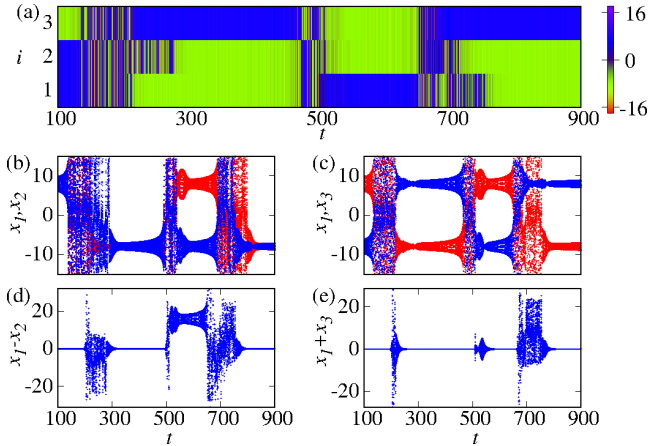}
\caption{(Color online) Intermittent chimera shown in $N = 3$ coupled Lorenz oscillators (Eq.~\ref{eq:system-eq1}) on a ring with coupling strength $\ep = 0.11$. 
The space time plot with variable $x$ for all oscillators is shown in (a). Time series of the oscillators are compared by plotting variables $x_1, x_2$ and  $x_1, x_3$ in (b) and (c) respectively. In subplots (d) and (e) we plot the difference $x_1-x_2$ and sum $x_1 + x_3$, respectively to show synchronization and anti-synchronization between the oscillators.}  
\label{fig8} 
\end{figure}

% Figure ------------------------------------------------------------------
\begin{figure}
\centering
\includegraphics[scale=.330,angle=270]{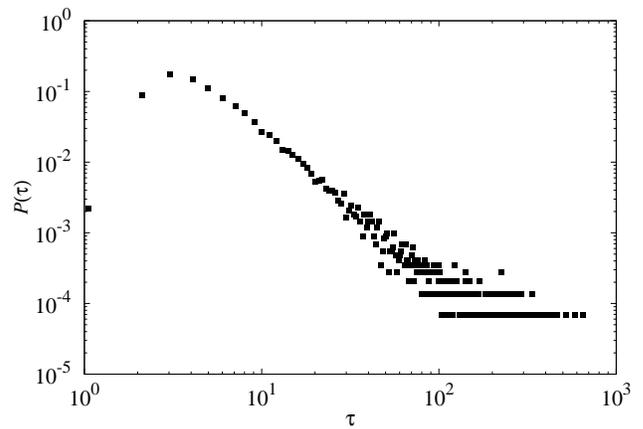}
\caption{Power law behavior of the distribution of interleaved intervals $\tau$ at $\epsilon=0.12$.}  
\label{fig9} 
\end{figure}

% Figure ------------------------------------------------------------------
\begin{figure}
\centering
\includegraphics[scale=0.33,angle=270]{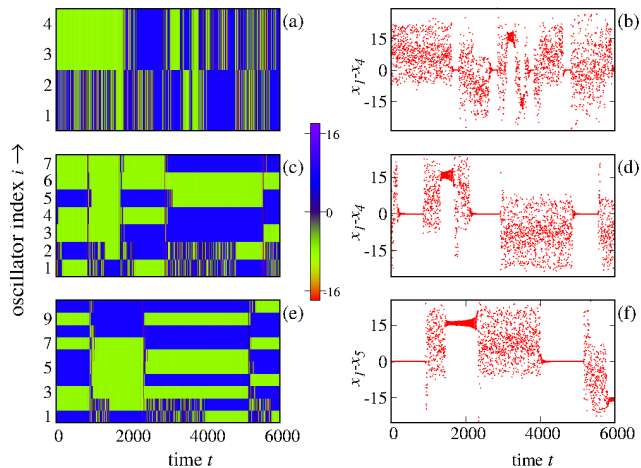}
\caption{(Color online) Intermittent dynamics in coupled Lorenz oscillator for different networks sizes. The results for $N=4$ (a, b), $7$ (c, d) and $10$(e, f) are plotted in each row at coupling strength  $\ep = 0.12$. The space time plots and difference variables indicating synchronization are plotted on the left and right panel respectively. The results confirm the existence of intermittent chimera states for larger networks.}  
\label{fig10} 
\end{figure}

We observe that for the coupling range $0.011<\ep<0.17$ trajectories  display intermittent synchronization or intermittent anti-synchronization. To avoid the effects of numerical artifacts on properties of intermittency \cite{Baker-1998}, we have performed simulations with double precision. As shown in Fig.~\ref{fig8}(a) at $\ep=0.11$, oscillator $1$ shows intermittent synchrony with oscillator $2$, for which the time series is plotted in Fig.~\ref{fig8}(b). Intermittent synchrony is evident from the time series of the difference of their $x$ variables namely $x_1-x_2$ in Fig.~\ref{fig8}(d). Similarly, anti-synchronization is observed between oscillators $1$ and $3$ for which the time series is plotted in Fig.~\ref{fig8}(c). Sum of these variables, $x_1+x_3$ is plotted in  Fig.~\ref{fig8}(e) which shows intervals of anti-synchrony and desynchrony. We explore the nature of intermittent synchronization/desynchronization following the procedure by Baker {\etal} \cite{Baker-1998}. We store the interleaved intervals of synchronized (anti-synchronized) dynamics and sort them according to the duration $\tau$ of the individual segments. The distribution of $\tau$ turns out to be a power law (see Fig.~\ref{fig9}) as expected for the case of intermittent dynamics~\cite{Hramov-2005}. 

We also observed such intermittent behavior for different network sizes. The results for $N=4,7,10$ oscillators on a ring are shown in Fig.~\ref{fig10}. The left panel of the figure shows the space-time plot. The time series of the difference variable $x_i-x_j$ are plotted in the right panel. These results verify the existence of intermittent synchrony/anti-synchrony in larger systems. The intermittent behavior of the oscillator appear due to the frustration introduced by neighbouring oscillators with different dynamical properties: we found that such intermittent synchrony and anti-synchrony is not observed for globally coupled oscillators. An oscillator in one attractor hops between different attractors depending upon the forcing from its neighbours.

The emergence of intermittent chimera states observed in our system can be explained as following. In a ring topology considered here, the dynamics of an oscillators depends on the signals received from its left and right neighbour. When the neighbouring oscillators asymptote to different attractors, the oscillator may show frustrated dynamics and pulled towards another manifold. We find, due to the such frustration and the complex manifold structures, oscillators' trajectories may show biased motion and may not even be contained within one attractor. For example, the oscillators on one attractor may stay in the vicinity of another attractor for relatively longer time, resulting in a biased motion shown in Fig.~\ref{fig2} (a, b). Further, the trajectories may also jump from one to another attractor (see Fig.~\ref{fig2} (c, d)). With increasing coupling strength, the frustration get stronger making these jumps more frequent (see Fig.~\ref{fig5}). This gives rise to intermittent dynamics observed in the system. Such intermittent jumps between different attractors also change the synchronized property of the system resulting the system to show {\it intermittent chimera} states.

\section{Intermittency in a network of non-locally coupled oscillators} \label{sec:nonlocal}

To substantiate our findings for smaller values of $N$, we consider an ensemble of nonlocally coupled oscillators for which the dynamical equations are given by:
 \begin{equation} \label{eq:nonlocal-ring}
 \begin{split}
 \dot{x}_i & =  \rho(y_i - x_i), \\
 \dot{y}_i & =  \gamma x_i - y_i - x_i z_i, \\
 \dot{z}_i & =  x_iy_i - \beta{z_i} + \frac{\ep}{2p}\sum_{j = (i - p)}^{(i + p)} (z_j-z_i),
\end{split}
\end{equation}
where, index $i=1,2,3, \cdots,N$, $N$ being the number of oscillators on the network. Each oscillator in the network is connected symmetrically with $2p$ nearest neighbours ($p$ to its left and $p$ to its right) with coupling strengt $\ep$. The coupling radius of the network is $r=p/N$~\cite{Abrams-2004}. For $p = 1$, oscillators are coupled only to one of its nearest neighbour in both sides (ring topology) and coupled scenario is referred to as the local with coupling radius $r = 1/N$. For global coupling, all the oscillators are connected to each other, \ie, $p = (N-1)/2$ or $ \approx N/2$ and coupling radius $r = (N-1)/2N \approx 0.5$ depending on whether $N$ is odd or even. The value of $r$ is between these limits refers to non-locally coupled scenarios~\cite{Dudkowski-2016,Maistrenko-2014}.
%

% Figure ------------------------------------------------------------------
\begin{figure}
\centerline{\includegraphics[scale=0.35,angle=270]{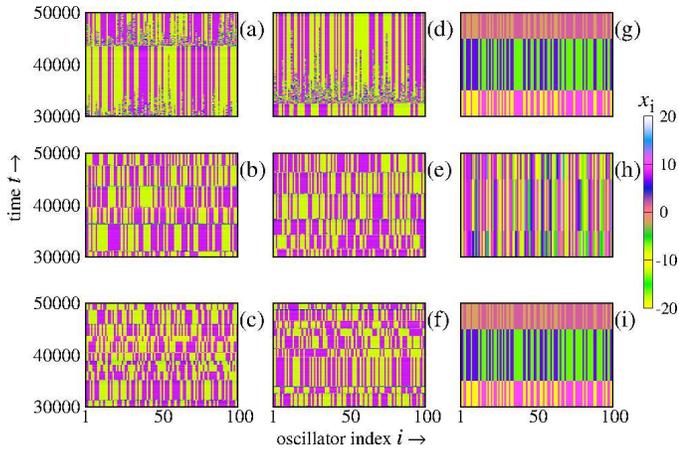}}
\caption{(Color online) Time evolution of space variables $x_i$ for a system of $N=100$ coupled oscillators (Eq.~\ref{eq:nonlocal-ring}) at different coupling strengths and coupling radii. The results for nonlocal coupling with coupling radii respectively $r = 0.1~(p=10)$ and $r = 0.35~(p = 35)$ are shown in left and middle panels. The right panel shows the results for global coupling with $r = 0.5~(p=50)$. The results for different coupling values are plotted in the top ($\ep=0.09$), middle $\ep=0.20$ and last row $\ep=0.35$. Intermittent behavior is observed for non-local coupling scenarios where the intermittent jumps become more frequent with increasing coupling strength. For globally coupled oscillators, intermittent behavior is not observed.}
\label{fig11}
\end{figure}

Fig.~\ref{fig11} represents time evolution with respect to space variables $x_i$. Left (Figs.~\ref{fig11}(a)-(c)) and right (Figs.~\ref{fig11}(d)-(f)) panels show plots for $N=100$ nonlocally coupled oscillators (Eq.~\ref{eq:nonlocal-ring}) for $p=10$ and $p=35$, respectively. In all the cases we observe that the dynamics of nonlocally coupled network exibits intermittent spatio-temporal chimera states. For smaller values of coupling strengths individual oscillator in the network spends long time in a particular attractor \ie,~ hopping around different attractors occurs slowly. Further, as the value of coupling parameter is increased, the time spends near an attractor by the individual entity present in the ensemble becomes shorter and hopping happens at very fast rate.

We have verified these cases for nearest neighbour couplings (ring) with large number of oscillators (say, $N=100$). We find that at smaller values of $\ep$, intermittent chimera states exist and at higher $\ep$ values coupled network left only with clusters of synchronized states where system dynamics evolve towards either $A_+$ or $A_-$ attractor. 

We also find that when coupling radius is increased to its optimum value, \ie, for global coupling, such intermittent behavior is not observed. This confirms that the intermittent behavior emerges as a result of frustration introduced by local and nonlocal coupling scenarios. 

\section{Summary}

In the present work, we study collective behaviour of coupled Lorenz oscillators near Hopf boundary. This system shows multistable behavior where the coexisting attractors have different dynamical properties. Due to such multistability, such systems are known to exhibit chimeric behavior in a globally coupled setting. Here, our motivation is to understand the effect of topology on such chimeric behavior. Specifically the effect of frustration, which is introduced in the system using ring topology with nearest neighbour and nonlocal interactions. We find that due to the introduction of frustration in this multistable system, the dynamics becomes intermittent. As a result of this intermittent behavior, the oscillators hop between synchronized and desynchronized motions, and we observe intermittent chimera states in the system. We also observe that this intermittent dynamics becomes more prominent as the coupling strength is increased. This suggests that the hopping of the oscillators depends on the interactions with its neighbours generating strong frustrations responsible for such behavior.

Using an example of $N=3$ oscillators on a ring, we show how the dynamics may exhibit biased behavior, namely, when the trajectory remains in the vicinity of another attractor for longer time, or exhibit intermittent jumps to other attractors. This intermittent behavior makes resulting dynamics of larger ensembles even richer where multiple oscillators may hop intermittently between different attractors. In such oscillator ensemble, the trajectories of the oscillators lie in a complex subspaces, namely, synchronized, anti-synchronized and desynchronized manifolds. The oscillators from the network may split or merge from the synchronized sub-population depending on the synchronization property of the attractor towards which it intermittently goes to. As a result of such hoping, oscillators may switch manifolds and the phenomenon of intermittent chimeras is observed in the system. The complexity of the coexisting attractors can be further explored by plotting the local exponential divergence plots which can be applied directly to the time series data. It is evident from the divergence plots  that the coexisting intermittent attractors are chaotic in nature. These attractors are responsible for the occurrence of intermittent synchronization/anti-synchronization for which the locking times show a power law decay.

The intermittent dynamics we observe here is due to the frustration created by its neighbours. This happens when the trajectories of neighbouring oscillators lie in different manifolds, \ie, they are moving on different attractors. The connected oscillator is entrained towards both of these motions creating the frustration, which consequently leads to the hopping between different manifolds and therefore intermittent dynamics shown by the oscillator. This condition is readily fulfilled in the ring topology, specially for local and non-local coupling strategies. However, for the globally coupled oscillators such intermittent behavior is not observed since all oscillators feel an averaged mean field effect due to all-to-all coupling. 

Intermittent chimera states have been observed in experimental systems with or without multistable dynamics, for example, waveguide resonators~\cite{Clerc-2017}, electrochemical systems~\cite{Schmidt-2014} and mechanical rotators~\cite{Olmi-2015}. In addition to these time continuous systems, existence of such states was recently reported in discrete-time systems of coupled map lattices~\cite{Singha-2020}. It is an open question whether such states can be observed in cellular automata as well. Chimera states have already been observed in various experimental set-ups including Josephson Junction arrays and coupled laser. Therefore, it would be interesting to explore intermittent behavior of multistability induced chimera states in such systems.  

\label{sec:conclusion}

%

%

%%%%%%%%%%%%%%%%%%%%%%%%%%%%%%%%%%%%%%%%%%%%%%%%%%%%%%%%%%
%%%%%%%%%%%%%%%%%%%%%%%%%%%%%%%%%%%%%%%%%%%%%%%%%%%%%%%%%%

\end{document}